\documentstyle[sprocl,epsfig]{article}
\newcommand{\lwig}{\mbox{\,\raisebox{.3ex}
    {$<$}$\!\!\!\!\!$\raisebox{-.9ex}{$\sim$}\,}}
\newcommand{\gwig}{\mbox{\,\raisebox{.3ex}
    {$>$}$\!\!\!\!\!$\raisebox{-.9ex}{$\sim$}}\,}

\def\Journal#1#2#3#4{{#1} {\bf #2}, #3 (#4)}

\def\NPB{{\em Nucl. Phys.} B}
\def\PLB{{\em Phys. Lett.}  B}

\def\PRD{{\em Phys. Rev.} D}

\def\EPJC{{\em Eur. Phys. J.} C} 

\def\be{\begin{equation}}
\def\ee{\end{equation}}
\def\bea{\begin{eqnarray}}
\def\eea{\end{eqnarray}}

\begin{document}

\title{{\vspace{-1.7cm}\normalsize\rightline{DESY
00-089}\rightline{hep-ph/0006215}}\vskip 0.3cm
    INSTANTONS IN DEEP-INELASTIC SCATTERING\footnote{Talk presented at the 
    8th International Workshop on Deep Inelastic Scattering (DIS\,2000), 
    Liverpool/UK, April 25-30, 2000; to be published in the Proceedings.}}

\author{\vspace{-0.5cm}A. RINGWALD AND F. SCHREMPP}

\address{Deutsches Elektronen-Synchrotron (DESY), Notkestrasse 85, 
D-22607 Hamburg, Germany\\E-mail: 
ringwald@mail.desy.de, fridger.schrempp@desy.de}

\maketitle\abstracts{In view of the new (preliminary) 
search results for instanton-induced events at HERA from the H1
collaboration, we present a brief 
discussion of (controllable) theoretical uncertainties, both in the
event topology and the calculated rate.}

Instantons~\cite{bpst,th} are a basic aspect of QCD. Being
non-perturbative fluctuations of the gauge fields, they induce hard
processes absent in conventional perturbation theory~\cite{th}.
Deep-inelastic scattering (DIS) at HERA offers a unique opportunity~\cite{rs1} 
to discover processes induced by instantons ($I$) through a sizable
rate, calculable within ``instanton-perturbation
theory"~\cite{mrs,rs2,rs-lat}, along 
with a characteristic final-state signature~\cite{rs1,cgrs}. Among the most
important features are a ``fireball''-like final state with a very 
high number of hadrons, including K-mesons and Lambda-hyperons, as 
well as a high total transverse energy. With the help of the Monte
Carlo generator QCDINS for $I$-induced events in
DIS~\cite{qcdins}, the experiments at HERA are actively 
searching for signatures of instantons in the hadronic final state. 

A more extensive introduction to $I$-induced events in DIS may be
found elsewhere~\cite{rs-rev}.  
In view of the new (preliminary) search results from the H1
collaboration~\cite{mikocki}, let us concentrate on a brief 
discussion of (controllable) theoretical uncertainties~\cite{pheno},
both in the event topology and the calculated rate.

\begin{figure}
\vspace{-0.7ex}
\begin{center}
\parbox{4.75cm}{\includegraphics*[width=4.75cm]{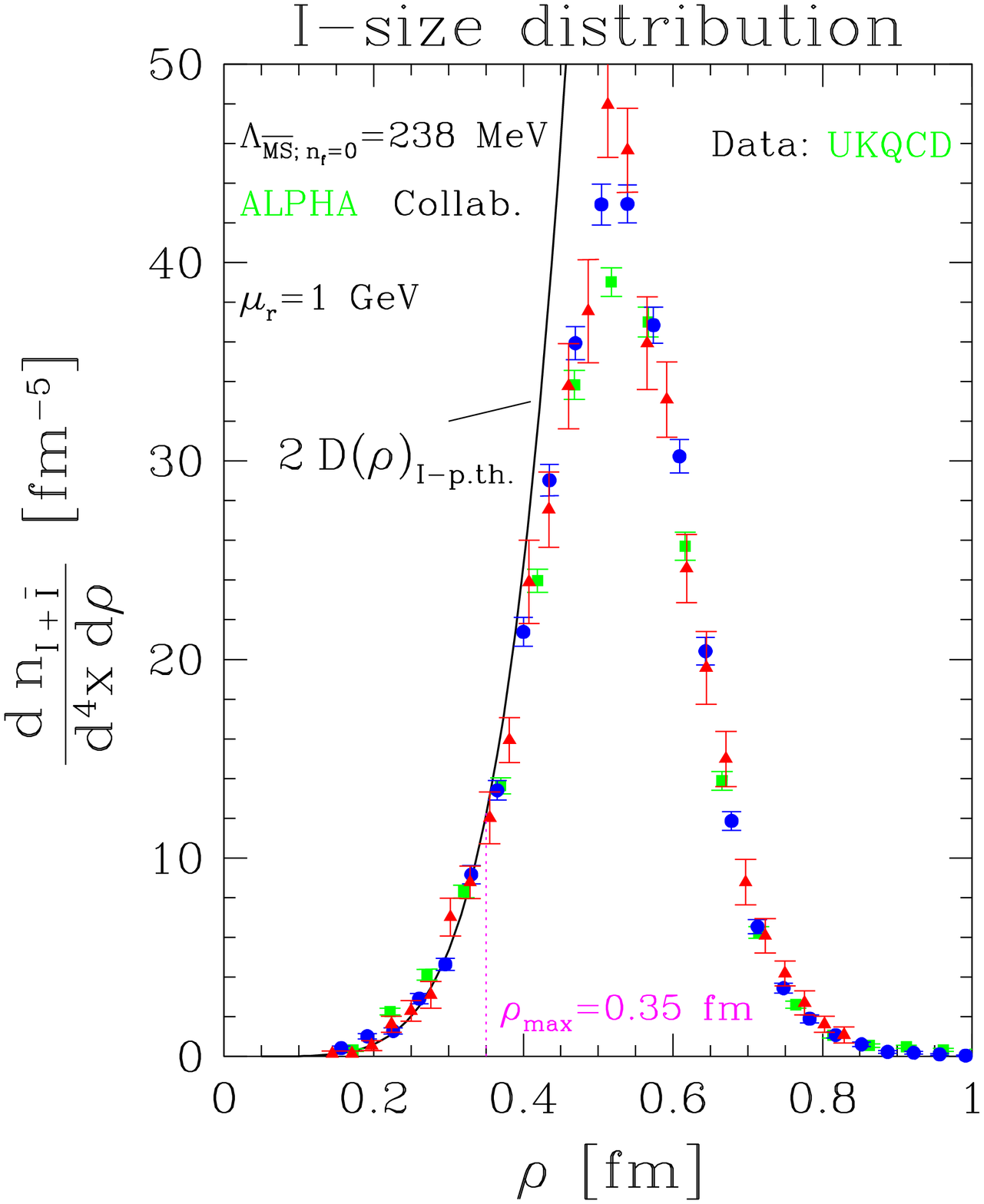}}
\hspace{0.5cm}
\parbox{4.75cm}{\includegraphics*[width=4.75cm]{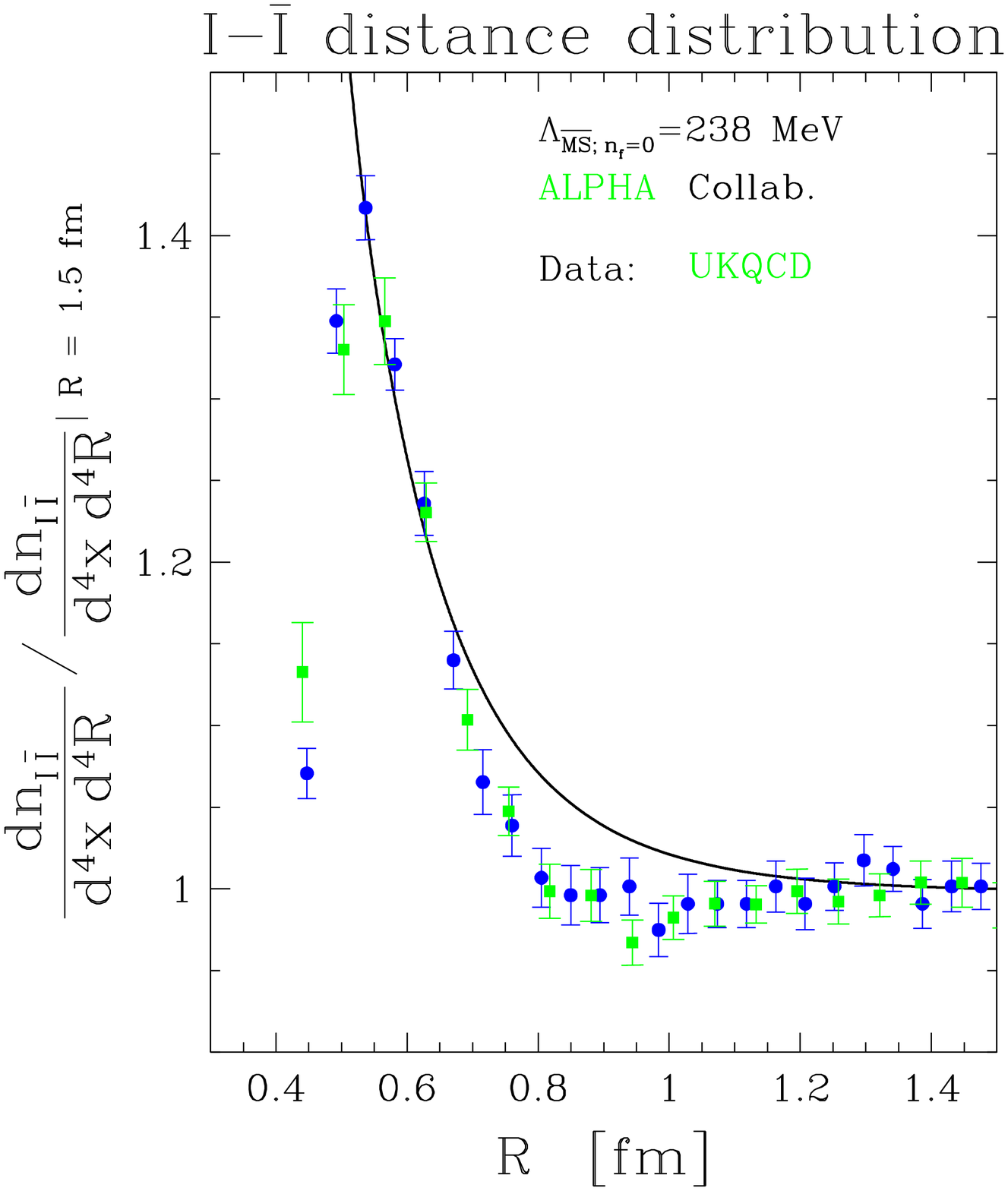}}
\caption[dum]{$I+\overline{I}$-size distribution (left) and normalized
$I\overline{I}$-distance distribution (right) compared to the
respective predictions from $I$-perturbation theory (solid
lines)~\cite{rs-lat}. \label{lattice}} 
\end{center}
\vspace{-0.4cm}
\end{figure}

\begin{figure}[b]
\vspace{-0.3cm}
\begin{center}
\includegraphics*[width=8.0cm]{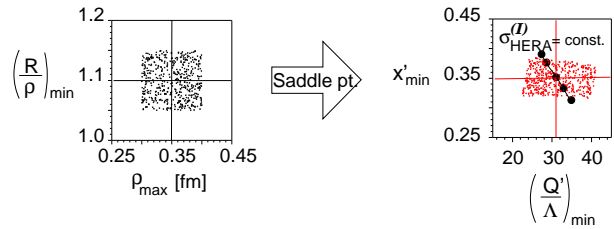}
\caption[dum]{One-to-one saddle-point translation of the restrictions
on the validity of $I$-perturbation theory from lattice ``data''
(Fig.~\ref{lattice}, Eq.~(\ref{rhoR}))  into
cuts on the associated momentum-space variables in DIS at
HERA. \label{momspace}} 
\end{center}
\end{figure}

Our predictions for $I$-induced events at HERA are based
on $I$-pertur\-bation theory~\cite{rs2}. 
The validity of the latter requires instantons of
small enough size $\rho\le\rho_{\rm max}$ along with a sufficiently
large separation $R/\rho\ge (R/\rho)_{\rm min}$ between them. Crucial
quantitative information on $(\rho_{\rm max},\ (R/\rho)_{\rm min})$ was
obtained~\cite{rs-lat} by  confronting the predictions of
$I$-perturbation theory with the ``data'' from a recent high-quality lattice
simulation (c.f. Fig.~\ref{lattice}). They lead us to conclude that
$I$-perturbation theory is quantitatively reliable for
\begin{equation}
 \rho\,\lwig\,\rho_{\rm max}=0.35\pm 0.05 {\rm\ fm}; \ \ 
 R/\rho\,\gwig\, (R/\rho)_{\rm min}=1.1\pm 0.05.
\label{rhoR}
\end{equation}
Let us emphasize that the $I$-induced cross section $\sigma^{(I)}_{\rm
HERA}$ depends strongly on the lattice observables in Fig.~\ref{lattice}
via a Fourier-type transformation. Therefore, it is
not too surprising that these constraints may be translated in a
one-to-one manner via a saddle-point relation  (c.f. Fig.~\ref{momspace})  
into {\it minimal cuts} on the conjugate Bjorken variables of the
$I$-subprocess $q^\prime g\stackrel{(I)}{\Rightarrow}
(2\,n_f-1)\,(q,\overline{q})+n_g\,g$,     
\begin{equation}
 \frac{Q^\prime}{\Lambda_{\rm \overline{MS}}}\ \gwig\  
 \frac{Q^\prime_{\rm min}}{\Lambda_{\rm \overline{MS}}}\
 \Leftrightarrow\ 
 \frac{1}{\Lambda_{\rm \overline{MS}}\,\rho_{\rm max}}\,;\ \ \ 
 x^\prime\ \gwig\  x^\prime_{\rm min}\ \Leftrightarrow\ 
 \left(\frac{R}{\rho}\right 
 )_{\rm min},
\end{equation}
on which $\sigma^{(I)}_{\rm HERA}$ crucially depends~\cite{rs2,rs-lat}.
The central values in Fig.~\ref{momspace} are part of the default cuts
in QCDINS 2.0~\cite{qcdins}. 
Note that it is very important to place the required cuts as close as
possible to the fiducial boundary, in order to retain a sufficiently
big event rate.

An important issue~\cite{pheno} is to study, how the remaining theoretical
uncertainties in these minimal 
$(Q^{\prime}/\Lambda_{\rm\overline{MS}}\,,
\,x^\prime)$ cut values (Fig.~\ref{momspace}) may affect the {\it
shape} of the six distributions of final
state observables~\cite{cgrs} used in the H1 analysis~\cite{mikocki}. 
\begin{figure}[t] 
\vspace{-0.5ex}
\begin{center}
\parbox{10cm}{\includegraphics*[angle=-90,width=10cm]{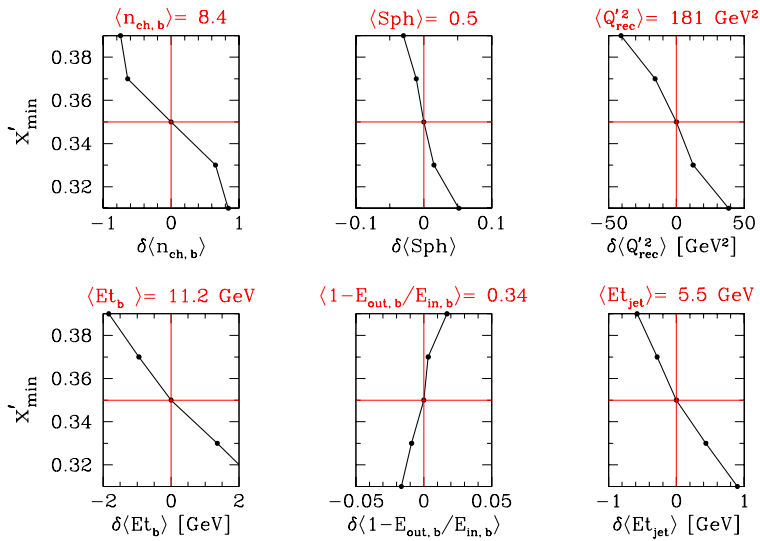}}
\caption[dum]{Shifts of the distribution averages of
observables used by H1~\cite{mikocki} for a  ``path'' through the
domain of allowed {\it minimal} cuts
$(Q^{\prime}_{\rm min}/\Lambda_{\rm\overline{MS}}\,,\,x^\prime_{\rm
min})$, corresponding to a constant $I$-induced rate
(c.f. Fig.~\ref{momspace}). $n_{\rm  
ch,\,b}$ denotes 
the number of charged particles within the $I$-band~\cite{rs1,cgrs},
$Sph$ the sphericity in the rest system of the particles not
associated with the current
jet, $Q^{\prime\,2}_{\rm rec}$ the reconstructed virtuality of the
quark $q^\prime$ entering the $I$-subprocess, $Et_{\rm b}$ the total
transverse energy within the  $I$-band, $1-E_{\rm out,\,b}/E_{\rm
in,\,b}$ the $Et$-weighted azimuthal isotropy and $Et_{\rm jet}$ the
transverse energy of the current jet, defined to be the jet with the
largest $Et$ found by the `cone'-algorithm. \label{obs} }
\end{center}
\end{figure}
Fig.~\ref{obs} illustrates the shifts on the averages of these
(peaked) distributions for a  ``path'' through the domain of allowed 
$(Q^{\prime}_{\rm min}/\Lambda_{\rm\overline{MS}}\,,\,x^\prime_{\rm min})$
pairs, corresponding to constant $\sigma^{(I)}_{\rm HERA}$
(c.f. Fig.~\ref{momspace}). Notably for the $Q^{\prime\,2}_{\rm rec}$
and $Et_b$ distributions, we  observe substantial shifts
of the averages (``peaks'') towards smaller values for increasing (i.e.
more reliable) $x^\prime_{\rm min}$. Note that we consider the lacking
experimental reconstruction~\cite{mikocki} of $x^\prime$ not a
serious problem as to the comparison with our predictions, since the 
lattice data for the $I\overline{I}$-distance distribution
(Fig.~\ref{lattice}, right) indicate an extremely strong suppression
of $I$-effects for small separation $R/\rho\lwig 1.1$, i.e. 
$x^\prime\lwig 0.35$.

A major uncertainty in the $I$-induced cross section arises from its
strong dependence on $\Lambda_{\rm\overline{MS}}$. In
Table~\ref{HERAcross}, we present our originally published value~\cite{rs2} of
$\sigma^{(I)}_{\rm HERA}$ based on $\alpha_s$ from PDG\cite{pdg96}\, ${}^\prime
96$ 
together with its update~\cite{qcdins} referring to $\alpha_s$ from
PDG\cite{pdg98}\, ${}^\prime 98$ for the default cuts used in QCDINS 2.0.    
\begin{table}[t]
\vspace{-0.5cm}
\caption[dum]{$I$-induced cross section for
HERA~\cite{rs2,qcdins} and its error from
$\delta\,\Lambda_{\rm\overline{MS}}$.\label{HERAcross}}  
\begin{center}
\begin{tabular}{|c||c|c||c|}\hline
\rule[-2.0ex]{0ex}{5ex}&PDG\cite{pdg96}\, ${}^\prime 96$ & {
PDG\cite{pdg98}\, ${}^\prime 98$}&{PDG\cite{pdg98}\, ${}^\prime 98$}\\\hline 
$\alpha_s(M_Z)$ & $0.113$ ({DIS}!) & $0.119\pm 0.002$ &
$0.119\pm 0.002$ \\\hline
$\Lambda^{(3)}_{\overline{\rm MS}}$ [MeV]& $260{{
+66}\atop{-64}}$&$346\,{{
+31}\atop{ -29}}  $ & $ 346\,{{ +31}\atop{ -29}}$\\\hline 
\rule[-3ex]{0ex}{5.5ex} & ${Q^\prime}\ge 8$ GeV&
$\frac{Q^\prime}{\Lambda^{(n_f)}_{\overline{\rm
MS}}}{\ge 30.8}$&$\frac{({Q,\,Q^\prime})}{\Lambda^{(n_f)}_{\overline{\rm
MS}}}{\ge 30.8}$\\\cline{2-4} 
\raisebox{2.0ex}[-2.0ex]{\bf\large Cuts} &\multicolumn{3}{c|}{\rule[-1.5ex]{0ex}{4.5ex}${
x^\prime\gwig  0.35},\ x_{\rm Bj}\ge 10^{-3},\ 0.1\le y_{\rm Bj}\le 0.9$}
\\\hline\hline
$\rule[-2ex]{0ex}{5ex}{\sigma^{ (I)}_{\mbox{ HERA}}}$ [pb] & $
126 {{ +300}\atop{ -100}}$ & ${ { 89}{{ -15}\atop{ +18}}}$ &
${{ 29}{{ -7.5}\atop{ +10}}}$\\\hline
\end{tabular}
\end{center}
\vspace{2ex}
\end{table}

\end{document}